\documentclass[aps,twocolumn,letterpaper,showpacs,superscriptaddress,10pt]{revtex4-1}
\usepackage[utf8]{inputenc}
\usepackage{graphicx,amsmath,color}
\usepackage{amssymb}
\usepackage{amsthm}
\usepackage{times}
\usepackage{bbm}
\usepackage[colorlinks,bookmarks=true,citecolor=blue,linkcolor=red,urlcolor=blue]{hyperref}

\newcommand{\jens}[1]{\textcolor{black}{#1}}
\newcommand{\je}[1]{\textcolor{black}{#1}}
\newcommand{\ejb}[1]{\textcolor{black}{#1}}
\newcommand{\eb}[1]{\textcolor{black}{#1}}

\newcommand{\jcb}[1]{\textcolor{black}{#1}}
\newcommand{\resub}[1]{\textcolor{black}{#1}}

\newcommand\cc{{\mathbbm{C}}}

\newcommand\ii{{\mathbbm{1}}}
\newcommand{\tr}{\text{Tr}}

\begin{document}

\title{Search for localized Wannier functions of topological band structures via compressed sensing}

\author{J.\ C.\ Budich}

\affiliation{Institute for Theoretical Physics, University of Innsbruck, 6020 Innsbruck, Austria}
\affiliation{Institute for Quantum Optics and Quantum Information, Austrian Academy of Sciences, 6020 Innsbruck, Austria}

\author{J.\ Eisert}
\affiliation{Dahlem Center for Complex Quantum Systems, Freie Universit\"at Berlin, Arnimallee 14, 14195 Berlin, Germany}

\author{E.\ J.\ Bergholtz}
\affiliation{Dahlem Center for Complex Quantum Systems, Freie Universit\"at Berlin, Arnimallee 14, 14195 Berlin, Germany}

\author{S. Diehl}
\affiliation{Institute for Theoretical Physics, University of Innsbruck, 6020 Innsbruck, Austria}
\affiliation{Institute for Quantum Optics and Quantum Information, Austrian Academy of Sciences, 6020 Innsbruck, Austria}

\author{P.\ Zoller}
\affiliation{Institute for Theoretical Physics, University of Innsbruck, 6020 Innsbruck, Austria}
\affiliation{Institute for Quantum Optics and Quantum Information, Austrian Academy of Sciences, 6020 Innsbruck, Austria}

\date{\today}
\begin{abstract}
\jcb{We investigate the interplay of band structure topology and localization properties of Wannier functions. To this end, we extend a} \eb{ recently proposed compressed sensing based paradigm for the search for maximally localized Wannier functions [Ozolins {\emph{et al.}}, PNAS {\bf{110}}, 18368 (2013)]. We
develop a practical toolbox that enables the search for maximally
localized Wannier functions which exactly obey the underlying physical
symmetries of a translationally invariant quantum lattice system under
investigation. Most saliently, this allows us to systematically identify the most localized representative of a
topological equivalence class of band structures, i.e., the most
localized set of Wannier functions that is adiabatically connected to
a generic initial representative. We also elaborate on the compressed
sensing scheme and find a particularly simple and efficient
implementation in which each step of the iteration is an $O(N \log N)$
algorithm in the number of lattice sites $N$. }
We present benchmark results on one-dimensional topological superconductors demonstrating the power of these tools. Furthermore, we employ our method to address the open question whether compact Wannier functions can exist for symmetry protected topological states like topological insulators in two dimensions. The existence of such functions would imply exact flat band models with finite range hopping. Here, we find
\je{numerical evidence} for the absence of such functions. 
We briefly discuss applications in dissipative state preparation and in devising variational sets of states for tensor network methods.
\end{abstract}
\maketitle

\section{Introduction and key results}
One of the most basic notions of condensed matter physcis is the quantum mechanical problem of a particle in a periodic potential. Yet, there are still quite fundamental questions relating to the physics of Bloch bands that have not been conclusively answered: How can optimally localized real space representations of band insulators in terms of Wannier functions (WFs) be found systematically and computationally efficiently? Under which circumstances can even compactly supported WFs exist for a given lattice Hamiltonian,
or at least for some representative of its topological equivalence class? These questions are of key importance not only for electronic band structure calculations within the single particle approximation, e.g., in the framework of density functional theory \cite{KohnSham}, but also for the dissipative preparation of topological band structures \cite{DiehlTopDiss} and their variational representation as a starting point for tensor network methods. In this work, we report substantial progress towards a comprehensive answer to these questions, \jcb{building on a {\it compressed sensing} (CS) based approach  to the problem of finding maximally localized WFs recently proposed by Ozolins et al. \ \cite{OszolinsCompressedModes,OszolinsTranslation}}.

\subsection{\je{Localized Wannier functions}}

The crucial optimization problem of finding maximally localized WFs $\lvert w_R^\alpha\rangle$ associated with a family of $n$ occupied Bloch vectors  $\lvert \psi_k^\alpha\rangle, \alpha=1,\ldots,n$\jens{,}
\je{and} $k\in \text{BZ}$ defined in the first Brillouin zone (BZ) has been subject of active research for many years \cite{VanderbiltReview}. The main difficulty is a local $U(n)$ gauge degree of freedom 
in reciprocal space acting on the Bloch functions as
\begin{align}
\lvert \psi_k^\alpha\rangle \je{\mapsto} \sum_{\beta=1}^n U_{\alpha\je{,}\beta}(k)\lvert \psi_k^\beta\rangle.
\label{eqn:gauge}
\end{align}
This redundancy in the definition of the Bloch functions renders the Wannier representation highly non-unique: A different gauge choice on the Bloch functions can modify the localization properties of the associated WFs which are defined as
\begin{align}
\lvert w_R^\alpha\rangle=\frac{V}{(2\pi)^{\je{d}}}\int_{\text{BZ}}\text{d}^{\je{d}}k\, \text{e}^{-ikR}\lvert \psi_k^\alpha\rangle,
\label{eqn:WFDef}
\end{align}
where $V$ is the volume of the primitive cell in real space and \je{$d$} is the spatial dimension of the crystal.

Interestingly, the search for maximally localized WFs is substantially influenced by topological aspects of the underlying band structure. The recent study of band structure topology has led to fundamental discoveries like topological insulators and superconductors \cite{HasanKaneXLReview} that have given a new twist to the basic physics of Bloch bands: Roughly speaking, the topology of insulating band structures measures the winding of the submanifold of occupied bands, represented by their projection $\mathcal P_k=\sum_{\alpha=1}^n \lvert \psi_k^\alpha\rangle\langle \psi_k^\alpha \rvert$, in the total space of all bands as a function of the lattice momentum $k$. The archetype of a topological invariant for band structures is the first Chern number 
\begin{align}
\mathcal C=\frac{i}{2\pi}\int_{\text{BZ}} \text{d}^2 k\, \text{Tr}\left( \mathcal P_k \left[(\partial_{k_x} \mathcal P_k),(\partial_{k_y} \mathcal P_k)\right]\right),
\label{eqn:Chern}
\end{align}
an integer quantized monopole charge associated with the gauge structure of the Bloch functions that distinguishes topologically inequivalent insulators in two spatial dimensions \cite{TKNN1982}. A non-vanishing monopole charge can be viewed as a fundamental obstruction to finding a global smooth gauge for the family of Bloch functions \cite{TKNN1982, Kohmoto1985}. However, it is precisely this analytical structure of the Bloch functions which determines the asymptotic decay of the associated WFs obtained by Fourier transform (cf.\ Eq.\ (\ref{eqn:WFDef})). This makes it intuitively plausible why a non-trivial band topology can have notable implications on the localization of WFs. Most prominently in this context, it is known that exponentially localized  Wannier functions exist if and only if the first Chern number 
is zero \cite{chernalgebraic, WannierChern, Matt}. In contrast, in one spatial dimension, Kohn could prove \cite{Kohn1959} that exponentially localized WFs always exist.

For so called symmetry protected topological states \cite{classification}, the situation is less simple. The topological nature of these band structures is rooted in the presence of a discrete symmetry, i.e., they are topologically distinct from an atomic insulator only if these underlying symmetries are maintained. Due to their vanishing Chern numbers, the existence of exponentially localized WFs is guaranteed for symmetry protected topological band structures. However, the possibility of representatives with even compactly  supported WFs is unknown for many symmetry protected states. A conclusive understanding of this issue is of particular interest since compactly supported WFs imply the existence of exact flat-band models with finite range hopping \cite{chernflat} and the possibility of dissipative analogs by virtue of local dissipation \cite{BardynTopDiss}.

A remarkably widely adopted and practically very useful approach to maximally localized WFs has been reported in Ref.\ \cite{MarzariVanderbilt1997}, see Ref.\ \cite{VanderbiltReview} for a recent review article. The guiding idea in Ref.\ \cite{MarzariVanderbilt1997} is to localize the WFs in real space by optimizing the gauge of the associated Bloch functions in reciprocal space based on a gradient search algorithm. Generically, this class of algorithms finds a local optimum that depends on the initial choice of gauge.

Very recently, a different paradigm for the construction of localized functions that approximately block-diagonalize a Hamilton operator has been formulated \cite{OszolinsCompressedModes}. This approach is rooted in the theory of CS \cite{Candes}, a contemporary branch of research at the interface between signal processing and fundamental information theory \cite{CSI}, which has also found applications in quantum theory  \cite{CS}.
In CS, the expected sparsity of a signal in some basis is employed for its exact reconstruction from significantly under-sampled measurement data, {without having to make use of the exact
sparsity pattern}. To this end, the sparsity of the signal is optimized under the constraint that it be compatible with the incomplete measurement data at hand. Translated to the spectral problem of a Hamiltonian, the analog of the incomplete measurement data is the ambiguity in the choice of basis functions that span a subspace associated with a certain energy range. Under the constraint of not involving basis states outside of this energy range, the sparsity of the basis functions in real space, i.e., their localization, is then optimized. First progress applying this program to the calculation of Wannier functions has been reported in Ref.\ \cite{OszolinsTranslation}.

\subsection{\je{Key results}}
In this work, we \jcb{extend} a CS based approach to the search for maximally localized WFs \jens{\cite{OszolinsCompressedModes,OszolinsTranslation} to study topological equivalence classes of band structures}. The physical motivation of our study is twofold: a comprehensive understanding of the interplay between band structure topology and localization properties of WFs at a fundamental level, and its impact on applications ranging from electronic band structure calculations over variational tensor network methods to dissipative state preparation.
To this end, \jcb{elaborating on the concepts introduced in Refs.\ \cite{OszolinsCompressedModes,OszolinsTranslation}}, we propose a numerically feasible and practical class of algorithms that are capable of manifestly maintaining the underlying physical symmetries of the band structure under investigation.  Most interestingly, this allows us to search for maximally localized representatives of a {\it topological equivalence class of band structures} via adiabatic continuity -- an unprecedented approach. The method exploring this possibility does not only search for a gauge of maximally localized WFs for a given Hamiltonian. Instead, the model Hamiltonian flows continuously within the symmetry class of band structures under consideration towards a topologically equivalent sweet-spot with compactly supported WFs. The starting point is in this case a set of Wannier functions of a generic representative of the topological state of interest.

The asymptotic scaling of each step of
\jcb{the present} iterative
method is $O(N\log(N))$, where $N$ is the number of lattice sites in the system. We argue that for each step this is up to constants the optimal effort:
any algorithm on such a translation invariant problem will at some point involve a fast Fourier transform which has the same scaling.  \jcb{This speedup compared to Ref.\ \cite{OszolinsTranslation}  is rooted in the use of a local orthogonality constraint imposed on the Bloch functions in reciprocal space that is equivalent to a non-local shift-orthogonality constraint on the WFs.} \jcb{Furthermore}, the \jcb{extended} algorithms proposed in this work are capable of exactly preserving the fundamental physical symmetries of the system under investigation.  \jcb{From a practical perspective,} this  can be of key importance to obtain physically meaningful results when constructing approximate Wannier functions for a given model. For example, if one is concerned with mean field superconductors in the {\it Bogoliubov de Gennes} (BdG) formulation, the fermionic algebra necessarily entails a formal {\it particle hole symmetry} (PHS) constraint;
its violation would lead to inherently unphysical results. \jcb{From a more fundamental perspective, the capability of manifestly maintaining the underlying symmetries at every iterative step opens us the way to study equivalence classes of topological bands structures instead of individual Hamiltonian representatives.} 

\begin{figure}[htp]
\centering
\includegraphics[width=0.76\columnwidth]{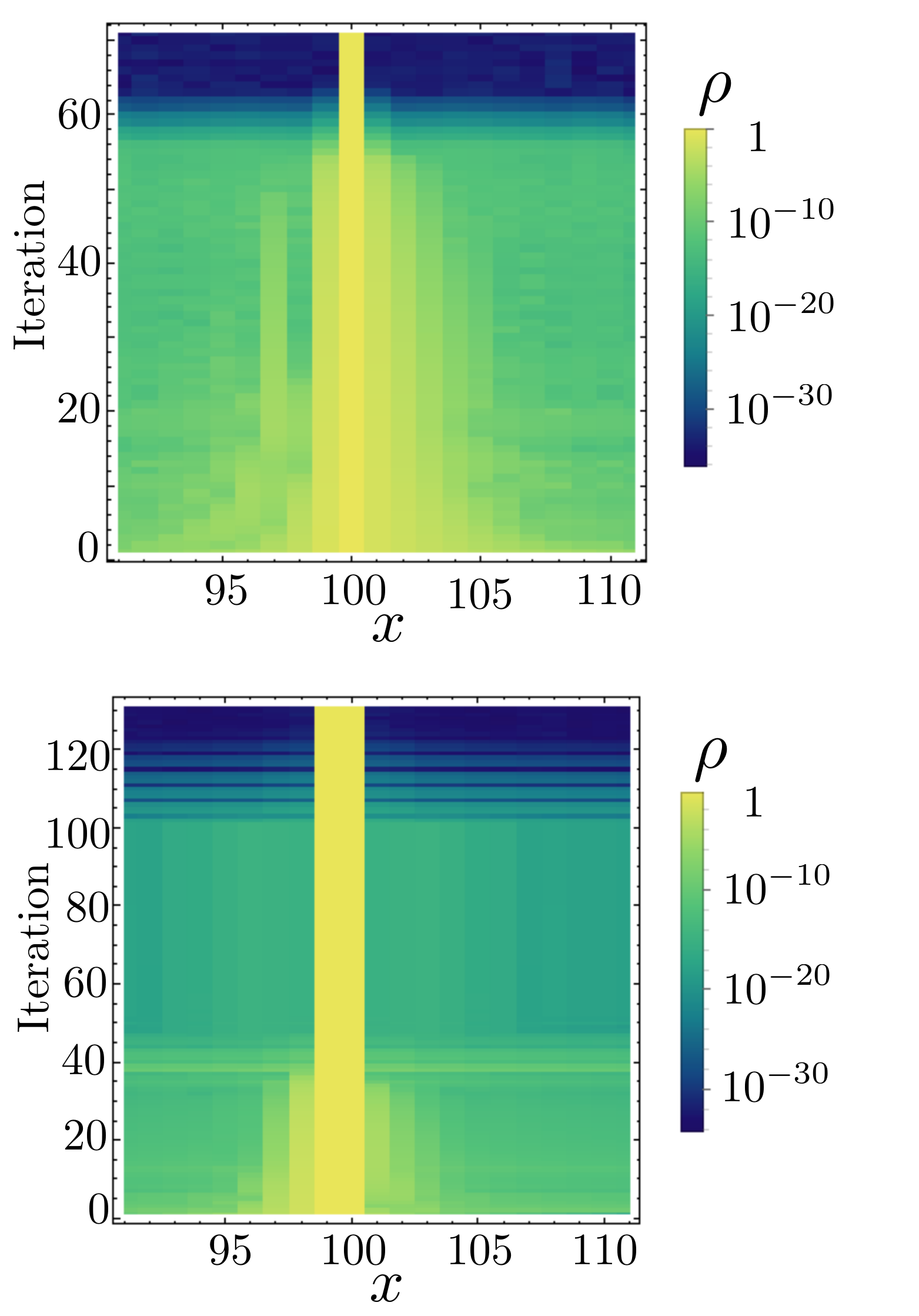}
\caption{(Color online) Evolution of the extent of Wannier functions under the adiabatic continuity algorithm for a trivial 1D superconductor  (upper panel) and a non-trivial 1D topological superconductor  (lower panel). In both cases, the most localized, compactly supported representatives of the respective phases are found \resub{, i.e., a WF localized on a single site (upper panel) and on two sites (lower panel), respectively}.  Plotted is the real space probability density $\rho_x$ (cf.\ Eq.\ \eqref{eq:rho}) on the horizontal $x$-axis with logarithmic color code from $10^{0}$ (yellow) to $10^{-30}$ (blue), and runtime increases on the vertical $t$-axis in units of ten iterative steps.  Initial Wannier functions obtain from the gauge constructed in Eq.\ (\ref{eqn:blochKitaev}) in Sec. \ref{sec:restop1D}. Parameters are $\mu =1.5, 2t=2\Delta=1$ and $\mu =0.3, 2t=2\Delta=1$ for upper and lower panel, respectively. The home cell of both WFs is $x=101$, with total length $L=200$ for both plots.}
\label{fig:movies1D}
\end{figure}

We present benchmark results for a one-dimensional (1D) topological superconductor (TSC)  \cite{Kitaev2001} demonstrating the efficiency of our method: Starting from a generic representative Hamiltonian of a 1D TSC, the algorithm converges towards a set of  WFs \resub{corresponding to a projection $\mathcal P_k$ onto an occupied band that obeys} the BdG PHS to high numerical accuracy. In the adiabatic continuity mode described above, our algorithm finds the maximally localized representative of the 1D TSC equivalence class, a state with compactly supported Wannier functions delocalized over two lattice sites. While for this particular state of matter, this ``sweet-spot'' point has been constructed analytically in Ref.\ \cite{Kitaev2001}, our search algorithm is capable of discovering it numerically starting from a generic model Hamiltonian represented by non-compact Wannier functions, as illustrated in Fig. \ref{fig:movies1D}.
For a topologically trivial starting point, the algorithm converges towards a set of atomic orbitals localized at a single lattice site -- the most localized representative of the trivial equivalence class. Finally, we give numerical evidence for the absence of compactly supported Wannier functions for {\it time reversal symmetry} (TRS) protected 2D topological insulators \cite{KaneMele2005a,KaneMele2005b,BHZ2006,koenig2007}: While our adiabatic search algorithm again converges to the WFs of an atomic insulator from a generic topologically trivial starting point, it does not find a compactly supported representative as soon as the initial Hamiltonian has gone through the phase transition to the topological insulator equivalence class. This indicates that there are no two-dimensional topological insulators with compact WFs.

{\emph{Outline. }}The remainder of this article is organized as follows. We define in Section \ref{sec:defopt} the search for maximally localized WFs associated with a given model Hamiltonian as an optimization problem subject to orthogonality and symmetry constraints. In Section \ref{sec:hamalg}, we \jcb{present} an efficient algorithm based on CS to numerically tackle this optimization problem. Numerical results for the  1D TSC are presented in Section \ref{sec:resham}. An algorithm which is not limited to a fixed model Hamiltonian but is designed for finding the most localized representative of a topological equivalence class of Hamiltonians is introduced in Section \ref{sec:algtop}. Benchmark results demonstrating the power of this tool are presented in Section \ref{sec:restop1D} and Section \ref{sec:restop2D}. 
Finally, we sum up our findings and give an outlook to possible applications in Section \ref{sec:conclusion}.

\section{Compact Wannier functions from sparsity optimization}
\label{sec:csham}
\subsection{Formulation of the optimization problem}
\label{sec:defopt}
The problem of calculating the electronic (fermionic) band structure of a crystal within the independent particle approximation can be viewed as the quantum mechanical problem of a single particle in a lattice-periodic potential. The spectrum of its solution consists of energy bands parameterized by a lattice momentum. Both eigenvalues and eigenfunctions are periodic with the reciprocal lattice and can hence be constrained to a representative unit cell of the reciprocal lattice called the first Brillouin zone (BZ). The eigenfunctions are so called Bloch states. For a given set of energy bands, WFs, i.e., localized functions in real space that are orthogonal to their own lattice translations (shift orthogonality)  can be obtained by Fourier transform of the Bloch states (cf.\ Eq.\ (\ref{eqn:WFDef})). In 1D, this problem has been addressed with methods from complex analysis by Kohn \cite{Kohn1959} showing that exponentially localized Wannier functions always exist. In higher spatial dimensions, topological obstructions can preclude the existence of exponentially localized WFs \cite{WannierChern}, e.g., due to a non-vanishing Chern number in 2D (cf.\ Eq.\ (\ref{eqn:Chern})). 

The work by Kohn \cite{Kohn1959}, as well as the majority of applications for band structure calculations \cite{VanderbiltReview}\je{,} focus on periodic problems in the spatial continuum. In practice, the continuous problem is often times not approximated by a straightforward discretization in real space but by deriving a so called tight binding model. The relevant degrees of freedom of such a model are then a finite number of orbitals per site of a discrete lattice with the periodicity of the crystalline potential. Our work is concerned with such lattice models within the independent particle approximation from the outset.

{\emph{Definitions. }}We consider a system with Hamiltonian $H$ on a hypercubic lattice with unit lattice constant and $N=L^d$~sites with periodic boundary conditions. Each lattice site hosts $m$~internal degrees of freedom (orbitals), $n$ bands are occupied. Our single particle wave functions are hence normalized vectors in $\cc^{m N}$, a set of Wannier functions is represented by a {matrix}
{$\psi\in \cc^{mN\times n}$}
with shift-orthonormal {columns}, i.e. $\psi^\dag T_j \psi=\ii \delta_{j,0} $ {for all} $ j\in \mathbb Z_L^d$, where $T_j$ performs a translation by the lattice vector $j\in \mathbb Z_L^d$. We denote the matrix elements by $\psi_{\nu, j; \alpha}$, where $\nu \in \{1,\ldots,m\}$,
$ j\in \mathbb Z_L^d$, and $\alpha \in \left\{1,\ldots,n\right\}$.  Among any set of shift orthogonal functions, a set of WFs associated with the $n$ occupied bands is 
distinguished by minimizing the quadratic energy functional
\begin{align}
\mathcal E[\psi] = \tr(\psi^\dag H \psi).
\label{eqn:evar}
\end{align}
While the Slater determinant forming the many body ground state characterized by its minimal energy expectation value of the insulating band structure is unique up to a global phase, the set of possible single particle WFs $\psi$ representing this ground state, i.e., minimizing $\mathcal E$ is highly non-unique. 
This is due to the local $U(n)$~gauge degree of freedom on the Bloch functions (cf.\ Eq.\ (\ref{eqn:gauge})).
Within this set, we would like to identify the representative where the probability density {$\rho_j^\alpha = \sum_\nu |\psi_{\nu , j; \alpha}|^2$} is most localized in real space. 
In the language of compressed sensing, localization is referred to as sparsity.
\jcb{As \jens{suggested} in Ref.\ \cite{OszolinsCompressedModes}, a $l_1$-norm regularization of the energy functional (\ref{eqn:evar}) is a convenient way to enforce the localization of the WFs. Concretely, as a measure for sparsity, we use} the vector
$l_1$-norm $\| \sqrt{\rho} \|{_{l_1}}=\sum_{j,\alpha}\lvert \sqrt{\rho_j^\alpha}\rvert$ of the square root of the probability density{, as a convex relaxation with more benign properties regarding continuity \resub{than} discrete measures like the rank.} 
For the WFs themselves, we define the $\rho$-norm as the $l_1$-norm of the square root of the associated probability density, i.e.,
\begin{align}\label{eq:rho}
\|\psi\|_\rho=\| \sqrt{\rho}\|_{l_1}.
\end{align}
A minimization with respect to the $\rho$-norm localizes the WFs only in real space and not in the internal degrees of freedom, as desired. The localization can be enforced by 
adding a term $ \| \psi\|_\rho/\xi$~ \jcb{to the  energy functional $\mathcal E$ \cite{OszolinsCompressedModes}}. The real parameter $\xi>0$~tunes the priority of the localisation respectively sparsity 
condition compared to the energy minimization condition. The optimization problem considered
is hence the minimization \jcb{of the $l_1$-regularized energy expectation  \cite{OszolinsCompressedModes}}
\begin{equation}
	 \mathcal E(\psi) + \frac{1}{\xi}\|\psi\|_\rho.
	 \label{eqn:l1reg}
\end{equation}
such that $\psi^\dagger T_j\psi=\ii \delta_{j,0}$. {The latter is a non-convex orthogonality constraint \cite{Yin}.}

Even if \jcb{the minimization of (\ref{eqn:l1reg})} will for finite $\xi$ in general produce approximations to  the WFs of the model characterized by $H$, we would like to make sure that the \resub{band structure defined by the resulting WFs preserves the} underlying physical symmetries of the problem exactly.  It is key to our algorithm that these symmetries can be exactly maintained.
Constraints that we will explicitly consider in this work are TRS $\mathcal T$, and PHS $\mathcal C$ (see Eq. (\ref{eqn:symconstraints}) for the corresponding constraints on the projection $\mathcal P_k$).  Generically, we denote the set of local symmetry constraints by $\mathcal S$. With these definitions, the problem of maximally localized WFs can {for each $\xi>0$} be concisely stated as the {$l_1$ regularized minimization problem}
\begin{align}
&\psi=\text{arg min}_\phi \left(\mathcal E(\phi)+\frac{1}{\xi} \|  \phi \|_\rho\right),\nonumber\\
&\text{subject to}\quad (\phi^\dag T_j \phi = {\ii }\delta_{j,0})~\text{and}~\mathcal S,
\label{eqn:minfunctional}
\end{align}
where arg gives the argument that minimizes the functional. The objective function is convex, while the symmetries \resub{and orthogonality constraints} give rise to 
quadratic equality
constraints.

\subsection{Compressed sensing based algorithm}
\label{sec:hamalg}

Convex $l_1$ regularized problems can be practically and efficiently solved using a number of methods. Here, we 
\jcb{use} a {\it split Bregman method} \cite{Bregman,Yin}, \jcb{which has been proposed to calculate maximally localized WFs in Refs.\ \cite{OszolinsCompressedModes, OszolinsTranslation}},
in a way that conveniently allows to include symmetries. The split Bregman method is related to the method of multipliers \cite{59}, which again can be connected to the alternating direction method of multipliers \cite{29}. Each step can then be implemented with as little as $O(N\log N)$ effort in the system size $N$.

The idea of a split Bregman iteration is to decompose the full optimization problem defined in Eq.\ (\ref{eqn:minfunctional}) into a set of coupled {subproblems} that can be solved exactly at every iterative step. We start from the simplest case without additional symmetries $\mathcal S$. In this case, our algorithm can be viewed as a numerically more efficient modification of the algorithms introduced in Refs.\ \cite{OszolinsCompressedModes, OszolinsTranslation}, adopted for {and generalized to}
a lattice Hamiltonian with internal degrees of freedom. We define the auxiliary variables $Q,R$ and associated noise terms $q,r$ that have the same dimension as the set of WFs $\psi\in
\cc^{mN\times n}$. During every step of the iteration, $\psi$ will optimise the energy functional $\mathcal E$ augmented by bilinear coupling terms (see step (i) below), $Q$ will be subject to a {\it soft thresholding procedure} stemming from the $\rho$-norm optimisation (see step (ii)), and $R$ will be subject to the shift-orthogonality constraint defining a proper set of WF (see step (iii)).
The noise terms $q$ and $r$ are incremented by the difference between $\psi$ and the auxiliary variables $Q$ and $R$, respectively  (see steps (iv)-(v)).  The algorithm in the absence of symmetries $\mathcal S$ then reads 
{as pseudocode}
\begin{align}
&\text{Initialize } \psi=Q=R,~q=r=0.  {\text{ While not converged do}}\nonumber \\
&\text{(i) } \psi\mapsto\text{arg}\min_{\psi} \left( \mathcal E[\psi]+\frac{\lambda}{2}\lVert  \psi-Q+q\rVert^2_F+\frac{\kappa}{2}\lVert  \psi-R+r\rVert^2_F\right),\nonumber\\
&\text{(ii) } Q\mapsto \text{arg}\min_{Q}\left( \resub{\frac{1}{\xi}}\|  Q\|_\rho+ \frac{\lambda}{2}\lVert  \psi-Q+q\rVert^2_F\right),\nonumber\\
&\text{(iii) } R\mapsto\text{arg}\min_R\frac{\kappa}{2}\lVert  \psi-R+r\rVert^2_F,~\text{s.t.}~\tilde R_k^\dag \tilde R_k=\frac{\ii}{L^{d/2}}~\forall k,\nonumber\\
&\text{(iv) } q\mapsto q+\psi-Q,\nonumber\\
&\text{(v) } r\mapsto r+\psi-R,
\end{align}
where {$\lVert \ejb{M}\lVert_F\ejb{=({\sum_{i,j}|M_{i,j}|^2}})^{1/2}$ denotes the Frobenius matrix norm of a matrix $M$,
and  $ \tilde R_k$~}the Fourier transform of $R$~at momentum $k$. $\lambda,\kappa,\xi>0$ are coupling constants. 
The way this problem is split in parts, the 
{subproblems} (i)-(iii) afford an explicit exact solution {not requiring any optimisation, given by}
\begin{align}
&\text{(i)}~ \psi=(2H+\lambda+\kappa)^{-1}(\kappa(R-r)+\lambda (Q-q)),\nonumber\\
&\text{(ii)}~ Q=\text{Shrink}\left(A,\frac{1}{\lambda \xi}\right),\nonumber\\
&\text{(iii)}~ \tilde R_k=\tilde B_k U\Lambda^{-\frac{1}{2}}U^\dag.
\label{eqn:exactmin}
\end{align} 
Here $A=\psi+q,~ B=\psi+r$, 
\begin{equation}
	\text{Shrink}(\je{b},\epsilon)= \frac{\je{b}}{\lvert \je{b}\rvert} \max(0,\lvert \je{b}\rvert-\epsilon)
\end{equation}	
is applied independently to each of the \resub{$m$-spinors} $B_j^\alpha$ associated with {the} Wannier function $\alpha$~evaluated at site $j$. 
{Also,}
\begin{equation}
	\tilde B_k^\dag \tilde B_k=U\Lambda U^\dag
\end{equation}
{with $U$ unitary and $\Lambda$ diagonal,} is {an eigenvalue}
decomposition of the positive Hermitian matrix $\tilde B_k^\dag \tilde B_k$. The orthogonality constraint  $\tilde R_k^\dag \tilde R_k=\ejb{\ii}/{L^{d/2}}~\forall k$ on the Bloch functions occurring in step (iii) is equivalent 
{with} the shift orthogonality {constraints} $R^\dag T_j R={\ii} \delta_{j,0}~\forall j$ 
on the Wannier functions. However, due to the local nature of the further, step (iii) can readily be solved exactly as explicitly done above, whereas the numerically less 
efficient method of Lagrange multipliers has been proposed in Ref.\ \cite{OszolinsTranslation} to enforce the latter non-local constraint in real space. 
{This is true even though it arises from a convex problem with a quadratic orthogonality constraint.}
More explicitly, the Fourier transform involved in the implementation used in the present work scales as $O(N\log N)$ if a fast Fourier algorithm is used.
Each step of the procedure is hence efficient. Rigorous convergence proofs for 
split Bregman methods are known for $l_1$-regularized convex problems \cite{Convergence}. Here, including the equality constraints, there is still
evidence that the entire method is efficient and convergent as well, {in line with the findings of Ref.\ \cite{Yin}.}

Step (iii) of the above algorithm solves the following problem: Given a set of wave functions $B$, it finds the closest possible (in {Frobenius} norm) 
set of proper shift orthogonal Wannier functions. Imposing additional local symmetry constraints $\mathcal S$~further complicates step (iii) of the above algorithm. From our numerical data presented below, it becomes clear that imposing constraints like PHS can be of key importance to obtain physically meaningful results. The simplest way to implement such symmetries is by considering the projection 
\begin{align}
\mathcal P_k=\sum_{\alpha=1}^n \tilde \psi_k^\alpha \tilde \psi_k^{\alpha \dag} 
\label{eqn:defpofk}
\end{align}
onto the occupied Bloch states at momentum $k$. Local symmetries will basically impose local constraints on this quantity, the only significant complication being the complex conjugation $K$ involved in anti-unitary constraints like TRS and PHS which connects $k$ and $-k$.  Explicitly, for TRS $\mathcal T$ and PHS $\mathcal C$, we get the constraints 
\begin{equation}
	\mathcal T \mathcal P_k \mathcal T^{-1}=\mathcal P_{-k},\,\,
	\mathcal C \mathcal P_k \mathcal C^{-1}=1-\mathcal P_{-k}, 
\label{eqn:symconstraints}	
\end{equation}
respectively.
With these definitions, we are ready to formulate a symmetry purification procedure augmenting step (iii). To this end, we follow (iii) to obtain the closest Bloch functions for half of the BZ and calculate $\mathcal P_k$. For the other half of the BZ, $\mathcal P_k$ is then obtained by symmetry conjugation \resub{by virtue of Eq. (\ref{eqn:symconstraints})}. The Bloch functions spanning $\mathcal P_k$ for this second half are obtained by projecting \resub{the Bloch functions from the previous iteration} $\tilde B_k$ onto $\mathcal P_k$ and again performing an orthogonalization based on a eigenvalue decomposition. By this continuous gauge prescription, we make sure that an input function $\tilde B_k$ that already obeys the given symmetry is unchanged by the purification procedure. This ensures that the algorithm can become stationary for the desired solution. The choice how the BZ is divided into two halves is {to some extent}
arbitrary. However, the fact that the Bloch basis in which we perform this purification and the real space basis in which the thresholding (ii) is performed are maximally incoherent bases 
{\cite{Candes}}
prevents systematic effects of this choice. For a unitary local symmetry, no constraint between $k$ and $-k$ is introduced and the symmetry purification can be done locally at every point in momentum space.

In summary, the core of our method consists of iteratively shrinking the spatial extent of the WFs by a soft thresholding prescription while reestablishing symmetry and orthogonality constraints on the associated \resub{projection $\mathcal P_k$} at every step. The localization and compact support of the WFs  is enforced directly in real space by virtue of $l_1$-norm optimization. Split orthogonality and symmetry constraints enforce the defining properties of the desired WFs. For a search limited to WFs of a fixed lattice model Hamiltonian, the subspace corresponding to a certain subset of bands and with that to a certain energy range is selected by minimizing a quadratic energy functional as proposed in Ref.\ \cite{OszolinsCompressedModes}. Hence, the CS approach does not require the knowledge of an initial set of WFs as a starting point. The converged trial functions are compactly supported well-defined Wannier functions by construction. Their degree of localization can be  tuned arbitrarily by a sparsity parameter $\xi$, with a tradeoff in controlling their quality in representing the given model Hamiltonian. 

\subsection{Results for the 1D TSC state}
\label{sec:resham}
As an interesting benchmark example, we consider the 1D TSC proposed by Kitaev in 2001 \cite{Kitaev2001} which is distinguished from a trivial superconductor by a topological $\mathbb Z_2$-invariant. The simplest representative of this state is a 1D lattice of spinless fermions  modelled by the Hamiltonian
\begin{align}
H_p =\sum_j \left(-t c_j^\dag c_{j+1} +\frac{\mu}{2} c_j^\dag c_j-\Delta c_j c_{j+1}+\text{h.c.}\right),
\end{align}
where $t$ is a real nearest neighbor hopping constant, $\mu$~models a chemical potential, $\Delta$~is a proximity induced $p$-wave pairing.
{The collection of} $c_j~(c_j^\dag)$ ~are the fermionic annihilation (creation) operators. Introducing the {collection of} Nambu spinors $\Psi_j=(c_j,c_j^\dag)^T$ and their Fourier transforms $\tilde \Psi_k=(\tilde c_k,\tilde c_{-k}^\dag)^T$, $H_p$ can be written in the BdG picture as
\begin{align}
H_p=\frac{1}{2}\int_{0}^{2\pi}\tilde \Psi_k^\dag d^i(k)\tau_i\tilde \Psi_k,
\label{eqn:KitaevBdG}
\end{align}
where $\tau_i$ are Pauli matrices in Nambu space \je{and}
\begin{align}
	d^1(k)&=0,\\
	d^2(k)&=-2 \Delta \sin(k),\\
	d^3(k)&=\mu -2t\cos(k).
\end{align}
For simplicity, we {consider the specific case} $2\Delta=2t=1$. As a function of $\mu$, $H_p$ is then in the topologically non-trivial phase for $\lvert \mu \rvert<1$, critical for $\lvert\mu\rvert=1$, and trivial otherwise. The description in terms of Nambu spinors implies a formal doubling of the degrees of freedom while keeping the number of physical degrees of freedom fixed. 
This redundancy \resub{is reflected} in an algebraic constraint on the BdG Hamiltonian that can be viewed as a formal PHS $\mathcal C=\tau_1 K$, where $K$ denotes complex conjugation.
The BdG Hamiltonian (\ref{eqn:KitaevBdG}) is formally equivalent to an insulating band structure with one occupied and one empty band. The projection $\mathcal P_k$ onto the occupied band can be expressed as
\begin{align}
\mathcal P_k=\frac{1}{2}(1-\hat d^i(k) \tau_i),
\label{eqn:ptwoband}
\end{align}
where {$\hat d(k)={d}(k)/{\lvert d(k)\rvert}$}.

\begin{figure}[htp]
\centering
\includegraphics[width=0.72\columnwidth]{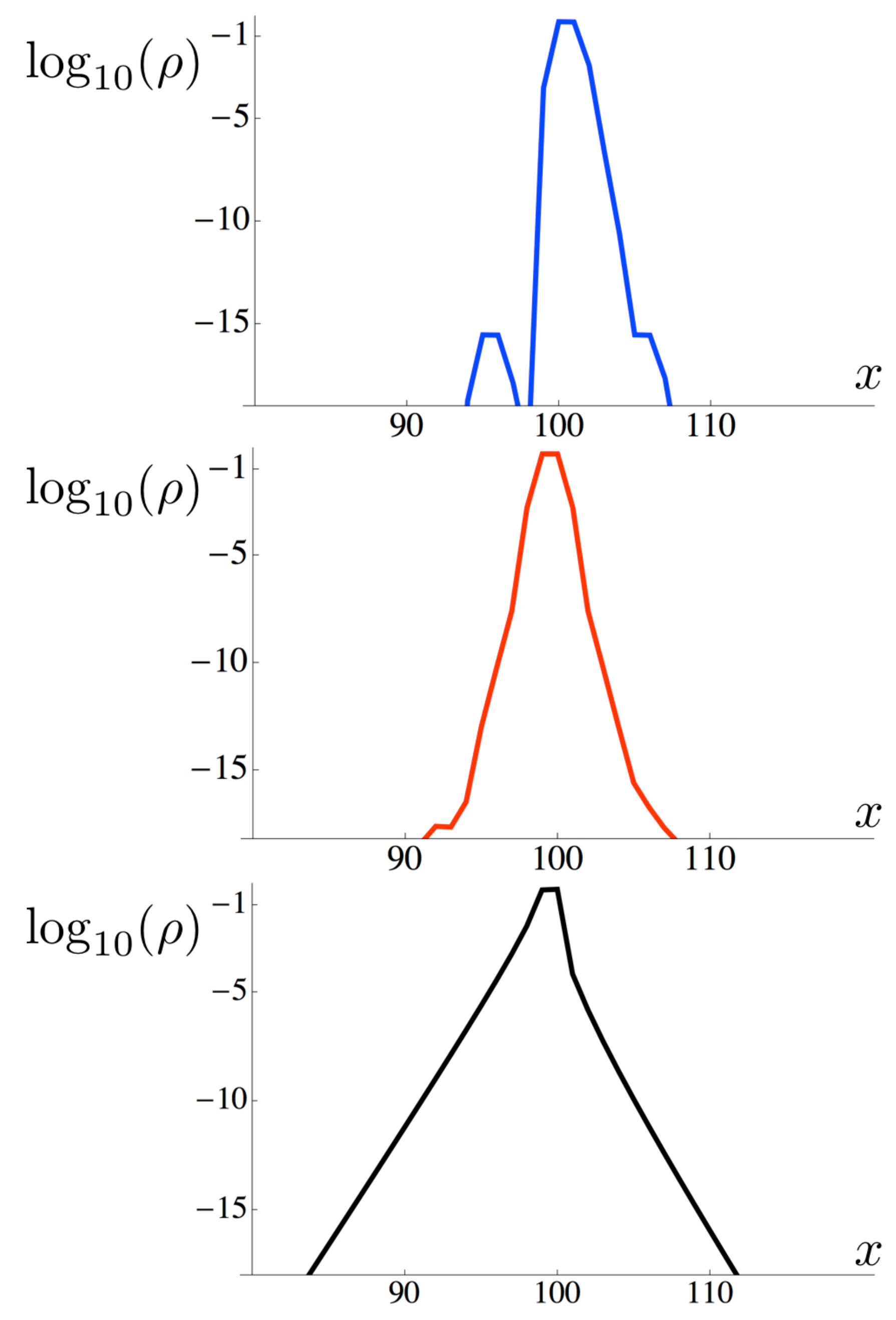}

\caption{(Color online) Logarithmic plots of the probability density $\rho$ of WFs with home cell  $x=101$ for a non-trivial 1D TSC with $\mu =0.3, 2t=2\Delta=1$ (see Section \ref{sec:resham} for definitions). \eb{Upper} panel: Result of algorithm without additional symmetries and coupling constants $\xi=10,r=50,\lambda=20$. Central panel: Result of algorithm with $\mathcal S=\left\{\text{PHS}\right\}$ and coupling constants $\xi=10,r=50,\lambda=20$. \eb{Lower} panel: WF from the gauge constructed in Eq.\ (\ref{eqn:blochKitaev}).  $L=200$ 
{has been} chosen
for all plots.}
\label{fig:hamcomparison}
\end{figure}

We now apply the algorithm introduced in Section \ref{sec:hamalg} to the toy model (\ref{eqn:KitaevBdG}). We first ignore the PHS constraint and apply the algorithm without further symmetries $\mathcal S$. For $\xi=10,r=50,\lambda=20,\mu=0.3,L=200$, it converges towards a set of compact WFs \resub{(see Fig.\ \ref{fig:hamcomparison} upper panel)} functions that minimize $\mathcal E$ to a relative accuracy of $1.8\je{\times}10^{-3}$ but that break PHS by as much as $2.0$ percent. The violation of PHS is measured by the deviation of $\lVert \mathcal P_k-(1-\mathcal C \mathcal P_{-k}\mathcal C^{-1})\rVert_F$ from zero \resub{(cf. Eq. (\ref{eqn:symconstraints}))} . Note that a set of WFs \resub{for which the associated projection $\mathcal P_k$} does not preserve the Nambu PHS $\mathcal C=\tau_x K$ cannot describe any physical superconductor. This demonstrates how important it is to manifestly maintain PHS here. In a next step, we apply the algorithm with $\mathcal S=\left\{\text{PHS}\right\}$ for the same parameters. It converges towards compactly supported WFs \resub{(see Fig.\ \ref{fig:hamcomparison} central panel)} which minimise $\mathcal E$ to a relative accuracy of $1.7\je{\times}10^{-3}$ and \resub{for which $\mathcal P_k$ preserves} PHS to $2.0\je{\times}10^{-8}$ accuracy within our numerical precision, i.e., six orders of magnitude better than without explicitly maintaining PHS. We show a logarithmic plot of the probability density $\rho$ of the converged WFs in Fig.\ \ref{fig:hamcomparison}. From these plots, it becomes clear why PHS is rather strongly broken if not explicitly maintained: The PHS breaking WFs \resub{(upper panel)} minimize the energy functional $\mathcal E$ to roughly the same accuracy but have a somewhat smaller $l_1$-norm at the expense of violating PHS. We compare the results of our algorithm to an analytically obtained WF \resub{(lower panel)} which has been computed from a smooth family of Bloch functions (see Eq.\ (\ref{eqn:blochKitaev}) below for its explicit construction), which clearly is much less localized (note the logarithmic scale).    

\section{Maximally localized representatives of topological equivalence classes}
\label{sec:cstop}
\subsection{Adiabatic continuity algorithm}
\label{sec:algtop}
In Section \ref{sec:hamalg}, we introduced an algorithm that is designed to find compactly supported WFs for a fixed model Hamiltonian. In this Section, we present a tool which searches for the most localized compactly supported WFs not only for a given Hamiltonian but {within} an entire topological equivalence class. Topological equivalence classes are the connected components of the set of all free Hamiltonians obeying certain local symmetries $\mathcal S$. In other words, starting from any Hamiltonian that preserves $\mathcal S$, its topological equivalence class is defined by all Hamiltonians that can be reached adiabatically, i.e., continuously without closing the band gap and without breaking $\mathcal S$.  We confine our attention to topological states relying on at least one symmetry constraint, i.e., states with zero Chern number. For states with non-zero Chern number, it is known that no representative with exponentially localized let alone compactly supported WFs can exist {\cite{Brouder}}.
 
The key idea of our adiabatic continuity algorithm is the following: 
Start from a set of WFs associated with a generic representative of a given topological equivalence class. Perform the split Bregman iteration introduced in Section \ref{sec:hamalg} with the crucial difference that the energy functional $\mathcal E$~is set to zero. That way the bias towards a particular model Hamiltonian is completely removed. However, the symmetries $\mathcal S$ are again restored at every step of the iteration and the $\rho$-norm optimization is continuous on a coarse grained scale controlled by the thresholding parameter ${1}/({\lambda \xi})$. Hence, the model Hamiltonian that the instantaneous WFs represent will flow continuously in the topological equivalence class of the Hamiltonian associated with the initial set of WFs. The only bias of this flow is the $\rho$-norm optimization, i.e., the localization of the WFs in real space 
by minimization of the ${{l_1}}$-norm of the square root of their probability density. Thus, the adiabatic continuity algorithm searches for the most localized representative of a given topological state of matter.
For the converged set of WFs, the corresponding Bloch functions are readily obtained by Fourier transform. From these Bloch functions, the projection onto the occupied bands $\mathcal P_k$ is straightforward to compute (see Eq. (\ref{eqn:PfromWanniers})). The generic flat band Hamiltonian $Q(k)=1-2\mathcal P_k$ then defines an explicit model Hamiltonian for the most localized representative of the topological equivalence class under investigation. 

\subsection{Maximally localized representatives in symmetry class D in one dimension}
\label{sec:restop1D}
To benchmark the adiabatic continuity algorithm, we would like to apply it to the 1D TSC model (\ref{eqn:KitaevBdG}) introduced in Section \ref{sec:resham}. 
{In the language of Ref.\ \cite{Altland}, this belongs to the  symmetry class D.}
For this model, the result of a perfect performance is clear: From a topologically trivial starting point, we would expect our algorithm to converge towards an ``atomic'' Wannier function which has support only on a single site. From Ref.\ \cite{Kitaev2001}, we also know the simplest representatives of the topologically nontrivial class, which are of the form $\lvert t\rvert =\lvert \Delta\rvert >0=\mu$. Such exactly dispersionless models are characterized by WFs \resub{corresponding to operators} of the form $w_j=(c_j+c_j^\dag-c_{j+1}+c_{j+1}^\dag)/2$ with compact support on only two sites around $j\in \left\{1,\ldots,L\right\}$. It is clear that no topologically non-trivial state can be represented by WFs with support on a single site, as this would preclude any momentum dependence of $\mathcal P_k$. We hence expect a set of WFs \resub{annihilated by operators} similar to $w_j$ as a result of our adiabatic search in the topologically non-trivial sector.

\begin{figure}
\includegraphics[width=0.7\columnwidth]{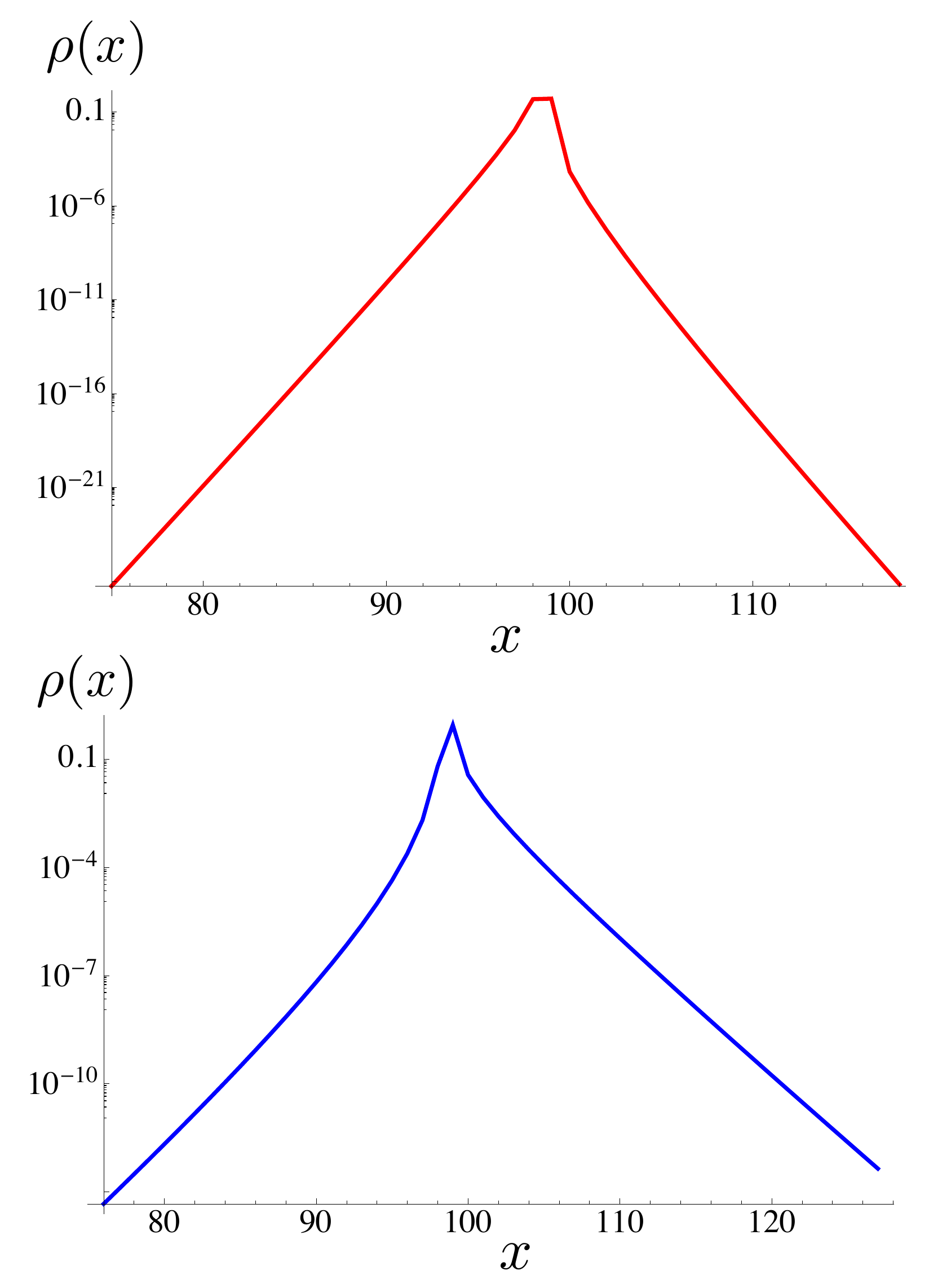}
\caption{(Color online) Logarithmic plot of the probability density $\rho$ of sets of Wannier functions from the gauge constructed in Eq.\ (\ref{eqn:blochKitaev}) for a trivial 1D superconductor with $\mu =1.5, 2t=2\Delta=1$ (lower panel) and a non-trivial 1D TSC with $\mu =0.3, 2t=2\Delta=1$ (upper panel). The home cell of both WFs is $x=101$. The linear tails demonstrate the asymptotic exponential decay. $L=200$ 
{is chosen}
for both plots.}
\label{fig:inWFLog}
\end{figure}

As a starting point we calculate a set of WFs from a family of Bloch functions representing the occupied band of $H_p$ for generic $\mu$. A global gauge defining a family of Bloch functions 
${k\mapsto \lvert u_-(k)\rangle}$ for the occupied BdG band can be constructed as
\begin{align}
\lvert u_-(k)\rangle = \frac{\mathcal P_k \lvert +x\rangle}{\lvert \mathcal P_k \lvert +x\rangle\rvert},
\label{eqn:blochKitaev}
\end{align}
where $\lvert +x\rangle=\tau_1  \lvert +x\rangle$ is a $\tau_1$ eigenvector. From Eq.\ (\ref{eqn:ptwoband}), it is easy to see that this gauge is regular for all $k$ since $d^1(k)=0$.
The initial WFs $\psi_0$~are then simply obtained by Fourier transform of the Bloch functions. Since $k\mapsto \lvert u_-(k)\rangle$ as resulting from Eq.\ (\ref{eqn:blochKitaev}) are $C^\infty$ functions, the corresponding Wannier functions are asymptotically bound to decay faster than every power law and exhibit in fact only exponential tails as verified in Fig.\ \ref{fig:inWFLog}. Our gauge choice turns out to be more efficient for the non-trivial WF which decays much more rapidly.

Using these functions as an input, the algorithm described in Section \ref{sec:algtop} indeed converges to the correct benchmark results in less than one minute on a regular desktop computer for a lattice of size $L=200$. In other words, our search algorithm numerically detects the ``sweet spot'' point with compactly supported WFs from Ref.\ \cite{Kitaev2001}, starting from a generic set of WFs representing some Hamiltonian with dispersive bands in the same topological equivalence class. Conversely, as soon as we tune $\mu$ over the topological quantum phase transition to a trivial 1D superconducting state, our search algorithm correctly finds an atomic WF representing the simplest possible trivial Hamiltonian. 

In Fig.\ \ref{fig:movies1D}, we visualize the performance of our algorithm with a logarithmic color plot of the probability density $\rho_x$ at lattice site $x$ as a function of the computation time $t$. The final WFs concur with the anticipated perfect benchmark results to impressive numerical precision.  

\subsection{Absence of compactly supported topological insulator WFs in symmetry class AII in 2D}
\label{sec:restop2D}
We would now like to turn our attention to time reversal symmetric 2D insulators, in  symmetry class AII \cite{Altland}. 
For states in symmetry class A with non-vanishing first Chern number, so called Chern insulators \cite{QAH}, only algebraically decaying WFs can be found. As a consequence, Chern insulators with exponentially localized or even compactly supported WFs cannot exist. 
However, the situation is less obvious for TRS protected topological insulators, a.k.a.\ quantum spin Hall (QSH) insulators \cite{KaneMele2005a,KaneMele2005b,BHZ2006,koenig2007}. The conceptually simplest representative of this topological equivalence class consists of two TRS conjugated copies of a Chern insulator with opposite odd Chern number, one copy for each spin block
{(cf.\ Ref.\ \cite{Matt})}. While the individual spin blocks have non-zero Chern number, the total set of occupied bands has zero Chern number as required by TRS. 
Hence, a smooth gauge mixing the two TRS conjugated blocks can be found for the Bloch functions \cite{Soluyanov2011}. 

{Here we} would like to consider a minimal model for a QSH insulator analogous to the one introduced in Ref.\ \cite{BHZ2006} which has $m=4$ degrees of freedom per site and $n=2$ occupied bands. The four degrees of freedom are labeled by the basis vectors $\vert e,\uparrow \rangle,\vert h,\uparrow \rangle,\vert e,\downarrow \rangle,\vert h,\downarrow \rangle$. We denote the $e-h$ pseudo spin by $\sigma$ and the real spin by $s$. The Bloch Hamiltonian of the spin up block reads as
\begin{align}
& h_{\uparrow}(k)=d_{\uparrow}^i(k)\sigma_i,\quad  d_{\uparrow}^1(k)=\sin(k_x),\nonumber\\
&d_{\uparrow}^2(k)=\sin(k_y),\quad d_{\uparrow}^3(k)=M-\cos(k_x)-\cos(k_y).
\label{eqn:BHZ}
\end{align}
The Hamiltonian of the TRS conjugated block is then {defined} by $h_\downarrow(k)=h^*_\uparrow(-k)$. This model is topologically nontrivial for $0<\lvert M\rvert < 2$ and trivial for $\lvert M\rvert >2$. The projection onto the occupied bands $\mathcal P_k$ can {for each $k$} be written as a sum of 
\begin{equation}
	\mathcal P_k^\uparrow =\frac{1}{2}\left(1-\hat d_\uparrow^i(k)\sigma_i\right)\otimes\lvert \uparrow\rangle\langle \uparrow \rvert
\end{equation}	
and 
\begin{equation}
	\mathcal P_k^\downarrow =\frac{1}{2}\left(1-\hat d_\downarrow^i(k)\sigma_i\right)\otimes\lvert \downarrow\rangle\langle \downarrow \rvert. 
\end{equation}	
A smooth gauge of Bloch functions ${k\mapsto \lvert u_i(k)\rangle}$, $i=1,2$, can be found in a generic way \cite{VanderbiltReview}. One first chooses a set of trial orbitals $\lvert \tau_i\rangle,~i=1,2${,} 
which are random linear combinations of the four basis orbitals. Projecting onto the occupied bands yields $\lvert \gamma_i(k)\rangle=\mathcal P_k \lvert \tau_i\rangle$. If the 
{family of} Gram {matrices} 
{with entries}
\begin{equation}
	S_{ij}(k)=\langle \gamma_i(k)\vert \gamma_j(k)\rangle
\end{equation}
 is regular for all $k$, smooth Bloch functions {defined as}
 \begin{equation}
 	{k\mapsto \lvert u_i(k) \rangle=S^{-{1}/{2}}_{j,i}(k) 
	\lvert \gamma_j\rangle }
\end{equation}
can be obtained. In practice, by trying a few random choices, a gauge for which ${\det(S(k) )\ge 10^{-2}}~\forall k$ can be readily found.
The associated WFs are then obtained by Fourier transform. \resub{Note that these WFs, while still spanning the same many-body state of occupied bands, individually break all symmetries present in Eq. (\ref{eqn:BHZ}) due to the random choice of $\tau_i$}.

\begin{figure*}
\centering
\includegraphics[width=.8\linewidth]{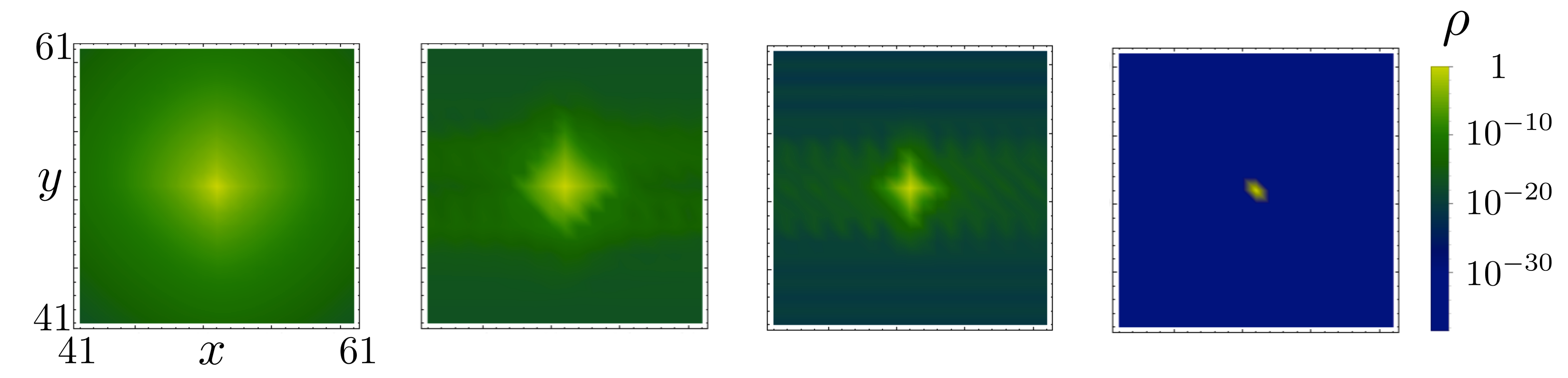}
\caption{(Color online) Logarithmic plot of the probability density $\rho$ of an initial Wannier function for the model Hamiltonian (\ref{eqn:BHZ}) for the topologically trivial mass parameter $M=2.5$ (leftmost panel). The home cell of the WFs is $(x,y)=(51,51)$, the size of the lattice is $101\times101$. Adiabatically deformed WF after 100, 500, and 727 (rightmost panel) iterations iterations respectively with $\xi=\kappa=\lambda=50$.}
\label{fig:movie2D}
\end{figure*}

We employ the above prescription to find exponentially decaying WFs both in the topologically trivial and nontrivial regime on a lattice of $N=101\times 101$ sites. These WFs are then used as starting points for the adiabatic continuity algorithm introduced in Section \ref{sec:algtop}. For WFs associated with topologically trivial insulators, i.e., $\lvert M\rvert>2$, our algorithm finds a set of atomic WFs representing the most localized topologically trivial insulator to impressive numerical accuracy (see Fig. \ref{fig:movie2D}). 
However, as soon as the initial set of WFs corresponds to a non-trivial QSH state, the algorithm does not find a compactly supported set of WFs. This result gives numerical evidence that a simple flat band model Hamiltonian with finite range hopping does not exist for the QSH state in contrast to the 1D TSC. The relation between flat band models with finite range hopping and compact WFs becomes clear from the following representation of the projection $\mathcal P_k$ onto the occupied bands at momentum $k$  in terms of Wannier functions $w^\alpha_0, \alpha=1,\ldots \je{,} n$ centered around the origin,
\begin{align}
\mathcal P_k = \sum_{\alpha=1}^n \sum_{r,r'}\text{e}^{ik(r-r')}w_0^\alpha(r) w_0^{\alpha \dag}(r').
\label{eqn:PfromWanniers}
\end{align}
An exact flat band Hamiltonian where all occupied states have energy $-\epsilon$ and all empty states have energy $+\epsilon$ is then immediately obtained as
\begin{align}
Q(k) = (+\epsilon) (1-\mathcal P_k) + (-\epsilon) \mathcal P_k=\epsilon\left(1-2\mathcal P_k\right).
\end{align}

To see if our findings are sensitive to the number of bands \resub{or to the spin rotation symmetry of Eq. (\ref{eqn:BHZ})}, we also applied the adiabatic continuity algorithm to a QSH model with 8 bands \resub{and spin mixing terms} which did not yield qualitatively different results.




\subsection{{Dissipative state preparation}}

The idea of dissipative state preparation \cite{DiehlStatePrep} in the context of topological states of matter \cite{DiehlTopDiss} relies, for pure and translation invariant target states, on the existence of a complete set of fermionic creation and annihilation operators $w_{i,\alpha},w_{i,\alpha}^\dag$ forming a Dirac algebra (the indices referring to sites and bands, respectively). In this case, the stationary state of a dissipative evolution described by a Lindblad master equation
\begin{eqnarray}
{\frac{\partial}{\partial t}}\rho = \kappa \sum_{i,\alpha} \left(w_{i,\alpha} \rho w^\dag_{i,\alpha}  - \tfrac{1}{2} \{w^\dag_{i,\alpha} w_{i,\alpha} ,\rho\}\right)
\end{eqnarray}
with damping rate {$\kappa>0$}, will be identical to the ground state of the dimensionless Hamiltonian 
\begin{equation}
	H = \sum_{i,\alpha}h_{i,\alpha},~  h_{i,\alpha} = w^\dag_{i,\alpha} w_{i,\alpha}, 
\end{equation}
with mutually commuting $h_{i,\alpha}$. In typical implementations of such a dissipative dynamics in the context of cold atomic systems, the Lindblad operators $w_{i,\alpha}$ generating the dissipative dynamics are quasi-local, i.e. have a compact support on the underlying lattice \cite{BardynTopDiss}. Our algorithm is precisely constructed to find such compactly supported operators $w_{i,\alpha}$, with the mutual commutativity of the associated $h_{i,\alpha}$ being granted by the shift orthogonality of the Wannier functions \resub{corresponding to the Lindblad operators $w_{i,\alpha}$}. Unlike the one dimensional case of the topologically nontrivial ground state of Kitaev's quantum wire, where a representative with compactly supported Wannier functions exists and is indeed found by our algorithm, our results in two dimensions imply the absence of an analogous situation in two dimensions.

\section{Conclusion and outlook}
\label{sec:conclusion}

In this work, we have presented a method to search for localized Wannier functions of free quantum lattice models which explicitly takes into account the symmetry of the problem. Most interestingly, we could extend the domain of this search algorithm from
individual model Hamiltonians to entire topological equivalence classes. This allows for a numerical detection of the most localized representative of a given topological state.  We did so by elaborating on a compressed sensing approach built upon Bregman split techniques, where the spatial locality takes the role of the sparsity of the problem (see Refs.\ \cite{OszolinsCompressedModes,OszolinsTranslation} ).
We close our presentation by providing some perspectives opened up by our present analysis, including a few particularly intriguing implications and applications of our new algorithm
\je{beyond the most widely known applications \cite{VanderbiltReview} of having localized Wannier functions available.}

\subsection{{Diagnostic tool of topological states}}

The possibility to identify localized Wannier functions not only for given model Hamiltonians, 
but also -- if the energy functional is set to zero along with $\xi\rightarrow 0$ --  maximally localized Wannier functions within entire
topological equivalence classes opens up another interesting application of our work: That of a {\it diagnostic tool}: Whenever it converges
\je{to a compactly supported Wannier function}, 
it identifies a "sweet spot" characterizing the topological class of the initial Hamiltonian itself rather than minimizing the energy of a certain model. The flow towards the atomic insulator and the topological flat band (Kitaev) superconductor, starting from generic states within the same topological phase provide striking examples of this. But the parameter $\xi>0$ can be freely chosen, reflecting the $l_1$-regularization in terms of compressed sensing. \je{In condensed matter terms, this parameter allows for a precise trade-off between locality and energy.}
This freedom gives rise to a useful ``knob'' to tune, and for applications in the context of e.g., ab initio band structure calculations, a finite $\xi$ is more appropriate.

\subsection{{Applications in devising tensor network methods}}

Thinking further about our algorithm as a flow in the renormalization group sense is likely to be fruitful also in the context of interacting as well as disordered systems. In fact our protocol bears some (non-accidental) resemblance with tensor network algorithms (quantum state renormalization methods) such as DMRG and TEBD in one dimension and PEPS and MERA more generally
\cite{R1,R2,R3}. 
More specifically, it seems that in order to simulate weakly interacting (and/or disordered) fermionic lattice models,
the efficiently localized Wannier functions which are still orthogonal appear to be a very suitable starting point for devising
variational sets of states, as real space operators remain short {ranged} (and close to diagonal) when projected to the pertinent electronic band. Most saliently, tensor network approaches augmented with an initial preferred basis selection based on our algorithm appear particularly promising in two-dimensional
approaches, where having a low bond dimension in PEPS approaches is critical for the highly costly (approximate) tensor network contraction.
More specifically, two approaches seem interesting: In a first, one takes a weakly interacting model and re-expresses the non-interacting part in the Wannier basis
found by the algorithm. If the Wannier functions are exactly localized, then the new Hamiltonian will still be local. This \je{can then serve as} 
an ansatz for a tensor network approach including 
interactions. In a second, one starts from a generalized mean field approach for the interacting model,
generates Wannier functions and then applies a variational
tensor network method.

\subsection{Symmetry breaking by truncation of exponential tails}
Finally, a fundamental question arises due to the apparent lack of \resub{compactly supported} Wannier functions for the quantum spin Hall phase, namely that of the importance of exponentially decaying tails. We have found that any truncation of the tail of the Wannier functions inevitably leads to the breaking of time-reversal symmetry at a corresponding rate. In fact, cutting exponential tails seems continuous, but the QSH phase can be left continuously by breaking TRS. Despite being a conceptual problem it may not be a practical one. In any solid-state realization of a finite size QSH insulator, there will be weak TRS breaking terms, yet the physical response can -- at least in principle -- be experimentally indistinguishable from that of a truly TRS invariant system. In this sense, even though the Wannier functions with compact support and formally do not represent a QSH phase, they may still be used for practical purposes.
Our algorithm provides a tool to systematically assess these questions. Yet these are merely a few of many intriguing directions, and we anticipate that our findings will inspire future research in diverse branches of physics, as well as in applied mathematics.\\

\jcb{\emph{Note added. A key result of the present paper is the use of {\emph{local}} orthogonality constraints on the Bloch functions. In this context, we note the recent arXiv submissions by Barekat \emph{et al.} \cite{Barekat1, Barekat2}. In Ref. \cite{Barekat1}, Barekat {\emph{et al.}} independently derive a similar algorithm with the same asymptotic scaling. In Ref. \cite{Barekat2}, the same authors use orthogonality constraints in terms of Bloch functions in the context of certain (topologically trivial) band structures. These papers do not address the maximally localized representatives of topological equivalence classes of band structures
which is the main focus of our present work.}}	

\section{Acknowledgements}

We would like to thank C.\ Krumnow and H.\ Wilming for discussions. \eb{We also thank V. Ozolins for helpful correspondence on Refs. \cite{OszolinsCompressedModes,OszolinsTranslation} and for making us aware of Ref. \cite{Barekat1}.}
{Support from the ERC \je{Synergy Grant} UQUAM and \je{Consolidator Grant TAQ}, the EU (RAQUEL, SIQS, \je{COST}), 
the BMBF (QuOReP), the START Grant No. Y 581-N16, the SFB FoQuS (FWF Project No. F4006- N16) and DFG's Emmy Noether program (BE 5233/1-1) is gratefully acknowledged.}

\bibliographystyle{apsrev}

\end{document}